\documentclass[american]{msml2020}
\usepackage[T1]{fontenc}
\usepackage[latin9]{inputenc}
\usepackage{amsmath}
\usepackage{amssymb}
\usepackage{graphicx}
\usepackage[dot]{bibtopic}

\makeatletter

\providecommand{\tabularnewline}{\\}

\usepackage{breqn}
\usepackage{algorithm}
\usepackage{algpseudocode}
\usepackage{bbold}
\usepackage{siunitx}
\usepackage{hyperref}
\DeclareSIUnit\steps{steps}
\renewcommand*{\vec}[1]{\mathbf{#1}}

\msmlauthor{
Luigi Sbail\`{o}\footnotemark[1] \and
Manuel Dibak\footnotemark[1] \\
\addr Freie Universität Berlin, Department of Mathematics and Computer Science, Arnimallee 6, 14195 Berlin, Germany \AND
Frank No\'{e} \Email{frank.noe@fu-berlin.de} \\
\addr Freie Universität Berlin, Department of Mathematics and Computer Science, Arnimallee 6, 14195 Berlin, Germany \\
\addr Freie Universität Berlin, Department of Physics, Arnimallee 6, 14195 Berlin, Germany \\
\addr Rice University, Department of Chemistry, Houston TX, 77005, United States 
}

\makeatother

\usepackage{babel}
\begin{document}
\title{Neural Mode Jump Monte Carlo}
\maketitle
\begin{abstract}
Markov chain Monte Carlo methods are a powerful tool for sampling
equilibrium configurations in complex systems. One problem these methods
often face is slow convergence over large energy barriers. In this
work, we propose a novel method which increases convergence in systems
composed of many metastable states. This method aims to connect metastable
regions directly using generative neural networks in order to propose
new configurations in the Markov chain and optimizes the acceptance
probability of large jumps between modes in configuration space. We
provide a comprehensive theory and demonstrate the method on example
systems.

\end{abstract}

\section{Introduction}

\renewcommand*{\footnoteseptext}{ }\renewcommand{\thefootnote}{\fnsymbol{footnote}}\footnotetext[1]{equal contribution}Markov
chain Monte Carlo (MCMC) methods are used to sample the equilibrium
distribution of systems whose probability distribution is otherwise
analytically intractable. An efficient MCMC generator proposes moves
that quickly decorrelate the samples while having a large acceptance
probability. A common choice are moves with a random displacement
in configuration space \cite{Metropolis}. As complex systems at equilibrium
visit only a small fraction of the whole configuration space, these
random displacements have to be very small to be accepted. However
small moves are only efficient at sampling local conformations of
the energy landscape, while crossing large energy barriers requires
a multitude of sampling steps. This problem is particularly evident
when the system is composed of many metastable states, where it is
often computationally infeasible to cross energy barriers multiple
times. This leads to a problem known as broken ergodicity, or quasi-ergodicity,
which implies that the probability to cross a energy barrier is so
low that simulations converge too slow to be practical.

In the last decades, many different methods have been developed to
circumvent this problem: One class of methods varies the temperature
during the sampling process as the crossing time over energy barriers
exponentially decreases with inverse temperatures. The two most widely
recognized methods in this class are simulated \cite{simulated_tempering,geyer1995annealing}
and parallel tempering \cite{geyer1991markov,hukushima1996exchange}
which operate on fixed set of temperatures. Simulated tempering randomly
changes the temperature of the sampler from a set of discrete temperatures
while remaining at equilibrium in an augmented temperature-configuration
space. In parallel tempering multiple simulations at different temperatures
are carried out in parallel and samples are randomly exchanged between
the different temperatures. These methods rely on a significant overlap
of the energetic distributions at different temperatures, therefore
the temperature range has to be chosen carefully.

A different class of methods biases the potential landscape in order
to enable transitions over energy barriers and recovers the unbiased
distribution by re-weighting. Metadynamics \cite{Laio} does this
in an iterative fashion, where the bias potential is increased in
areas where the system resides a long time thus pushing the system
out of metastable states. Recent development suggest the usage of
deep learning to find an optimal bias potential \cite{zhan2019:targeted}.
Umbrella sampling \cite{Torrie} runs several sampling iterations
with bias potentials placed along a pre-defined coordinate and thus
pushes the system from one end to the other of this reaction coordinate.

A novel method are Boltzmann Generators \cite{noe2019boltzmann},
which uses deep learning in order to learn to draw unbiased samples
from a target distribution $p_{X}(\mathbf{x})\propto\exp(-u(\mathbf{x}))$
by combining an exact probability generator such as a normalizing
Flow \cite{rezende2015:variational,dinh:2016density} with reweighting
\cite{noe2019boltzmann,albergo2010:flow,nicoli2019:asymptotically}.

Another recently developed approach \cite{Stern,NCMC_chodera,NCMC_roux}
constructs reversible moves between equilibrium states as a collection
of small out-of-equilbrium trajectories. This approach also depends
on the path connecting the equilibrium states, and a system specific
protocol to generate the candidate state must be designed.

Smart Darting Monte Carlo \cite{smart_darting,smart_darting_effectiveness}
is a promising method that alternates local and long range moves from
one region of the configuration space to another that is arbitrary
far. These moves are attempted between small spheres around local
minima. In high dimensions however the fraction of the spheres to
the total volume becomes vanishingly small and therefore finding a
sphere by random exploration becomes very unlikely. This problem is
circumvented in ConfJump \cite{ConfJump} by finding the closest energy
minimum, and attempting long range moves by translation to another
energy minimum.

The generation of long range moves is challenging when the energy
landscape is rough, since the potential energy surface in the region
surrounding local minima can drastically change between the different
minima. In this case, using trivial translation as long range moves
would most likely cause large energy differences and is likely to
be rejected. Instead, a specific bijective function pairing points
in order to keep the energy difference small needs to be employed.
However, constructing such bijection manually would require detailed
knowledge of the system and is practically impossible in multi-dimensional
systems.

Recent advances in the field of machine learning have permitted to
deal with problems that were not solvable with a sole human understanding,
and, more specifically, deep neural networks (DNNs) are an ideal tool
to facilitate the construction of a bijective function. DNNs have
already been employed to construct MCMC moves. Current methods use
DNNs to approximate the distribution and thus speeding up sampling
\cite{Self_learning_MC}, projecting onto high probability manifolds
\cite{habib2018auxiliary} or use a latent space representation in
order to propose moves \cite{noe2019boltzmann,albergo2010:flow}.

Two recent methods are using reversible network architectures in order
to improve Hamiltonian Monte Carlo (HMC): \cite{song:2017a-nice-mc}
propose steps by applying a volume preserving flow to the augmented
configuration space. \cite{levy2018generalizing} augment the leapfrog
algorithm commonly used in HMC with DNNs and thus alter the classical
path of the system while relying on forces. Both are trained for sampling
efficiency in a unsupervised fashion and therefore rely on random
exploration of configuration space in order to find metastable states.

In this paper, we present neural mode jump Monte Carlo (Neural MJMC),
a novel method to efficiently sample the equilibrium distribution
of complex many-body systems with unbiased Markov chains. In this
scheme neural networks are trained to propose \textquotedbl neural\textquotedbl{}
moves that directly connect different metastable states. These proposals
do not try to approximate a classical path between start and endpoint
which gives them the freedom to connect regions in phase space that
are arbitrarily far apart. The method requires a prior knowledge of
the position of the metastable states in configuration space, which
could e.g. be obtained from x-ray scattering experiments. Local displacements
and neural moves are randomly alternated in a combined scheme to accelerate
the convergence rate of Markov chains. Configurations from different
metastable states are used to train the networks, which are optimized
to produce high acceptance probability moves. Local exploration ensures
ergodicity of the scheme, while neural moves accelerate convergence
to equilibrium, realizing an accurate and deep exploration of the
configuration space.

\section{Theory}

A sufficient condition to ensure that a Markov chain asymptotically
samples the equilibrium distribution is ergodicity and detailed balance.
Given the system in a configuration $\vec{x}$ a new state $\vec{y}$
is added to the chain with a transition probability $p(\vec{x}\rightarrow\vec{y})$.
The transition probability is defined to satisfy the condition of
detailed balance

\begin{equation}
\pi(\vec{x})\thinspace p(\vec{x}\rightarrow\vec{y})=\pi(\vec{y})\thinspace p(\vec{y}\rightarrow\vec{x}),
\end{equation}
where $\pi(\vec{x})$ is the equilibrium distribution. In the Metropolis-Hastings
algorithm \cite{Metropolis,Hastings}, the transition probability
is decomposed in two logical steps: firstly, a new configuration $\vec{y}$
is drawn from a proposal density $p_{\mathrm{\mathrm{prop}}}(\vec{x}\rightarrow\vec{y})$,
then the new state is accepted with an acceptance probability $p_{\mathrm{acc}}(\vec{x}\rightarrow\vec{y})$.
If the transition is accepted, the new state $\vec{y}$ is added to
the Markov chain, otherwise the previous state $\vec{x}$ is added
to the Markov chain.

In Neural MJMC, additionally the proposal probability is split into
two steps: firstly, a proposal density is selected from a pre-defined
list of proposal densities on the current state $\vec{x}$, then a
new state $\vec{y}$ is drawn from the extracted proposal density.
Proposal densities are distinguished between local proposals and neural
proposals, where local proposals generate local moves through random
displacement, as already proposed in the Metropolis scheme, and neural
proposals connect different metastable states.

Let us assume that the configuration space $\Omega$ is decomposed
into a number of non overlapping subsets called cores $\{\Omega_{\alpha}\}_{\alpha\leq N}\subset\Omega$,
with $\cup_{\alpha}\Omega_{\alpha}=\Omega$, each representing one
of the $N$ metastable states. We define the neural proposal $K_{\alpha\beta}$,
as the density that proposes transitions from the core $\Omega_{\alpha}$
to the core $\Omega_{\beta}$. Assuming the system in the state $\vec{x}\in\Omega_{\alpha}$,
the probability to extract the neural proposal $K_{\alpha\beta}$
is $p_{\alpha\beta}(\vec{x})$. Once $K_{\alpha\beta}$ has been selected,
a state $\vec{y}\in\Omega_{\beta}$ is drawn from the selection probability
$\thinspace p_{prop}^{\alpha\beta}(\vec{x}\rightarrow\vec{y})$.

A neural proposal $K_{\alpha\beta}$ can only be selected within the
core $\Omega_{\alpha}$ and with constant probability $p_{\alpha\beta}(\vec{x})=p_{\alpha\beta}\chi_{\Omega_{\alpha}}(\vec{x})$,
where $\chi_{\Omega}(\vec{x})$ denotes the characteristic function.
We assume that each pair of states $(\alpha,\beta)$ is only connected
by one neural proposal $K_{\alpha\beta}$ and that there exists an
inverse proposal $K_{\beta\alpha}$ connecting $\beta$ with $\alpha$.
Under these assumptions, a proposed move starting in $\Omega_{\alpha}$
with selected neural proposal $K_{\alpha\beta}$ fulfills detailed
balance if it is accepted with probability

\begin{equation}
p_{acc}^{\alpha\beta}(\vec{x}\rightarrow\vec{y})=\min\left\{ 1,\frac{\pi(\vec{y})\thinspace p_{\beta\alpha}\thinspace p_{prop}^{\beta\alpha}(\vec{y}\rightarrow\vec{x})}{\pi(\vec{x})\thinspace p_{\alpha\beta}\thinspace p_{prop}^{\alpha\beta}(\vec{x}\rightarrow\vec{y})}\right\} .
\end{equation}
We parameterize the neural proposal $K_{\alpha\beta}$ and its inverse
$K_{\beta\alpha}$ connecting the cores $\Omega_{\alpha}$ and $\Omega_{\beta}$,
as a bijective function $\mu_{\alpha\beta}(\cdot)$ pairing the states
defined in the two cores, i.e $\vec{y}=\mu_{\alpha\beta}(\vec{x}$),
$\mu_{\alpha\beta}^{-1}(\vec{y})=\vec{x}$, $\forall\vec{x}\in\Omega_{\alpha}$,
where $\vec{x}\in\Omega_{\alpha}$ , $\vec{y}\in\Omega_{\beta}$.
Thus for each pair of different cores $(\Omega_{\alpha},\Omega_{\beta})$
a bijective function $\mu_{\alpha\beta}(\cdot)$ is defined. The probability
distribution of neural proposals is then represented with Dirac delta
distributions and the acceptance specifies to

\begin{equation}
p_{acc}^{\alpha\beta}(\vec{x}\rightarrow\vec{y})=\min\left\{ 1,\frac{\pi(\vec{y})\thinspace p_{\alpha\beta}\thinspace\delta(\vec{x}-\mu_{\alpha\beta}^{-1}(\vec{y}))}{\pi(\vec{x})\thinspace p_{\beta\alpha}\thinspace\delta(\vec{y}-\mu_{\alpha\beta}(\vec{x}))}\right\} .
\end{equation}
Using the change of variable formula in the Dirac distribution $\delta(\vec{x}-\mu_{\alpha\beta}^{-1}(\vec{y}))=\left|\det J(\mu_{\alpha\beta}(\vec{x}))\right|\delta(\vec{y}-\mu_{\alpha\beta}(\vec{x}))$,
the acceptance probability for neural moves can be simplified to

\begin{equation}
p_{acc}^{\alpha\beta}(\vec{x}\rightarrow\vec{y})=\min\left\{ 1,\frac{\pi(\vec{y})\thinspace p_{\alpha\beta}}{\pi(\vec{x})\thinspace p_{\beta\alpha}}\left|\det J(\mu_{\alpha\beta}(\vec{x}))\right|\right\} .\label{eq:acceptance_neural}
\end{equation}
In case that the local proposal ($\alpha=\beta$) is selected, the
inverse move is only possible with another local proposal $K_{\alpha\alpha}$.
Note that a local move may leave the current core and the proposal
probability for the inverse move might change. Thus the acceptance
probability for a local move reduces to

\begin{equation}
p_{acc}^{\alpha\alpha}(\vec{x}\rightarrow\vec{y})=\min\left\{ 1,\frac{\pi(\vec{y})\sum_{\beta}\chi_{\Omega_{\beta}}p_{\beta\beta}}{\pi(\vec{x})p_{\alpha\alpha}}\right\} .\label{eq:acceptance_local}
\end{equation}
In order to ensure ergodicity there needs to be a finite probability
of selecting the local proposal in all cores. In Algorithm \ref{sampling_algorithm},
we summarize the Neural MJMC sampling scheme.

\begin{algorithm2e}
\DontPrintSemicolon
\caption{Neural MJMC sampling scheme}
\SetKwInOut{Input}{input}
\Input{$l_{s}=[~]: \text{empty list for samples}$ \\
$\left\{p_{\alpha\beta}\right\}: \text{proposal selection probabilities}$\\
$\left\{\mu_{\alpha\beta}\right\}$: \text{proposal densities}\\
$\vec{x} \gets \vec{x}_0: \text{starting point of sampling}$\\
$N_{\textrm{iterations}}: \text{number of generated samples}$\\
$\sigma_{\textrm{local}}: \text{standard deviation of local moves}$
}
\While {$i \leq N_{\textrm{iterations}}$}{
	draw proposal density $K_{\alpha\beta}$ from $\left\{p_{\alpha\beta}\right\}$\;
	\uIf(\tcp*[f]{\text{propose local move}}){$\alpha =\beta$}
	{
		$\vec{w} \gets \text{sample from } \mathcal{N}(0,\mathbb{1}) $\;
		$\vec{y} \gets \vec{x} + \vec{w} \cdot \sigma_{\text{local}}$\;
		$p_{\text{acc}} \gets p_{\mathrm{acc}}^{\alpha\alpha}(\vec{x}\to\vec{y})$ (Eq. \ref{eq:acceptance_local})\;
	}
	\uElse(\tcp*[f]{\text{propose neural move}})
	{
		$\vec y = \mu_{\alpha\beta}(\vec x)$\;
		$p_{\text{acc}} \gets p_{\mathrm{acc}}^{\alpha\beta}(\vec{x}\to\vec{y})$ (Eq. \ref{eq:acceptance_neural})\;
	}
	\If{$r\sim \mathcal{U} (0,1) < p_{\text{acc}}$}{ $\vec{x} \gets \vec{y}$ }
	$l_{\text{s}}\text{.append }(\vec{x})$\;
	$i \gets i+1$\;
}
\label{sampling_algorithm}
\end{algorithm2e}

\subsection{Optimal proposal density}

In order to achieve fast decorrelation of the Markov chain the neural
proposal functions $\mu_{\alpha\beta}$ should maximize the acceptance
in both directions. This is quantified by maximizing the expected
log probability that the moves proposed by $\mu_{\alpha\beta}$ are
accepted in both directions. Using Jensen's inequality we find
\begin{multline}
\max_{\mu_{\alpha\beta}}\log\mathbb{E}_{\vec{x}\sim\Omega_{\alpha}}\left[p_{\mathrm{pacc}}(\vec{x}\to\vec{y})p_{\mathrm{pacc}}(\vec{y}\to\vec{x})\right]\geq\max\mathbb{E}\left\{ \log\left[p_{\mathrm{pacc}}(\vec{x}\to\vec{y})p_{\mathrm{pacc}}(\vec{y}\to\vec{x})\right]\right\} \\
=\max_{\mu_{\alpha\beta}}\mathbb{E}\left[\min\left(0,\log f\right)+\min(0,-\log f)\right]=\max_{\mu_{\alpha\beta}}\mathbb{E}\left[\min(\log f,-\log f)\right]=\max_{\mu_{\alpha\beta}}\mathbb{E}\left[-\left|\log f\right|\right],
\end{multline}
where $f=\frac{\pi(\vec{y})\thinspace p_{\alpha\beta}}{\pi(\vec{x})\thinspace p_{\beta\alpha}}\left|\det J(\mu_{\alpha\beta}(\vec{x}))\right|$.
Using the stationary distribution in the canonical ensemble $\pi(\vec{x})\propto\exp(-\beta V(\vec{x}))$,
with the thermal energy $\beta^{-1}=k_{B}T$, the potential energy
$V(\vec{x})$ of the system under consideration and assuming that
$\mu_{\alpha\beta}$ is a bijection between the cores $(\alpha,\beta)$,
we can rewrite the equation above to find

\begin{equation}
\min_{\mu_{\alpha\beta}}\mathbb{E}\left[\beta\left|\Delta V_{\alpha\beta}(\vec{x})+k_{B}T\log\left|\det J_{\mu_{\alpha\beta}}(x)\right|+\Delta R_{\alpha\beta}\right|\right],\label{eq:loss_acceptance}
\end{equation}
with the potential difference $\Delta V_{\alpha\beta}(\vec{x}):=V(\vec{x})-V(\mu_{\alpha\beta}(\vec{x}))$
and the log selection ratio $\Delta R_{\alpha\beta}:=-k_{B}T\log p_{\alpha\beta}/p_{\beta\alpha}$.
Note that the term inside the modulus is equivalent to the Kulbach-Leibler
divergence between the transformed distribution $\mu_{\alpha\beta}(\Omega_{\alpha})$
and the target distribution $\Omega_{\beta}$ as found in \cite{noe2019boltzmann}.

We can interpret this result in a physically meaningful manner by
applying the triangular inequality $\mathbb{E}\left[-\left|\log f\right|\right]\geq-\left|\mathbb{E}\left[\log f\right]\right|$,
identifying $\Delta S=-k_{B}\mathbb{E}\left[\log\left|\det J_{\mu_{\alpha\beta}}(x)\right|\right]$
as the change of entropy (see Appendix A for details) and $\Delta U=\mathbb{E}\left[\Delta V_{\alpha\beta}(\vec{x})\right]$
as the change of internal energy under the transformation $\mu_{\alpha\beta}(\vec{x})$.
We observe that the expected $\log$ acceptance is lower bound by
the absolute change in free energy $\Delta F=\Delta U-T\Delta S$
under the transformation $\mu_{ij}(\cdot)$ divided by thermal energy

\begin{equation}
\mathbb{E}\left\{ \log\left[p_{\mathrm{pacc}}(\vec{x}\to\vec{y})p_{\mathrm{pacc}}(\vec{y}\to\vec{x})\right]\right\} \geq-\beta\left|\Delta F+\Delta R_{ij}\right|.\label{eq:Delta_F_minimum-1}
\end{equation}
This result shows that we can use the freedom in the proposal selection
ratio in order to maximize the bi-directional acceptance.

\section{Neural network architecture}

\begin{figure}
\includegraphics[width=0.45\linewidth]{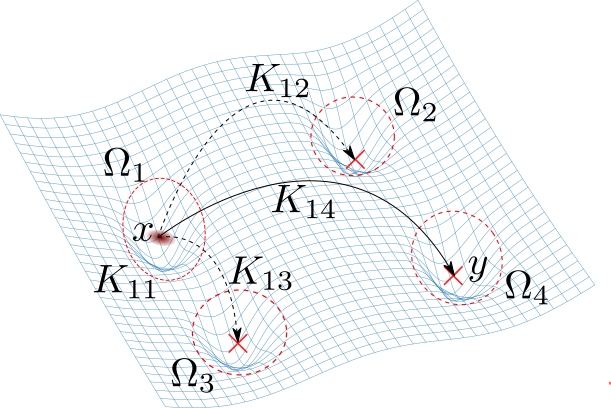}\includegraphics[width=0.45\linewidth]{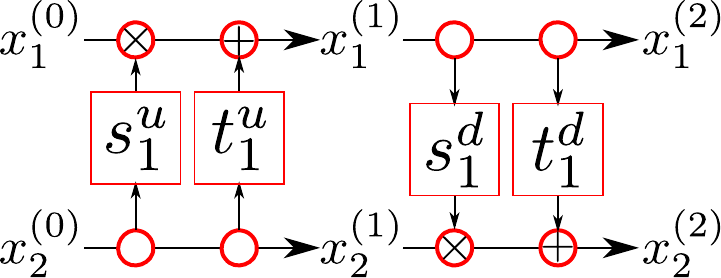}

\caption{\textbf{Left:} Schematic figure of Neural MJMC scheme. Given configuration
$\vec{x}$ in core $\Omega_{\alpha}$ there are three neural and one
local proposals available, as denoted by arrows. One of these is selected
and a new state $\vec{y}$ is proposed. \textbf{Right:} Architecture
of the RNVP networks that are used as reversible networks for the
examples in this paper. The input vector $\vec{x}$ is separated into
two disjoint sets of coordinates $\vec{x}_{1},\vec{x}_{2}$, and at
each iteration one subset undergoes a nonlinear transformation and
is multiplied and added to the other subset. The transformation can
easily be inverted.}

\label{Fig:rnvp}
\end{figure}

As a neural moves relies on an exactly invertible function with a
computationally feasible Jacobian, this must also be reflected in
the choice of the neural network architecture. Recently, several such
structures have been proposed \cite{song:2017a-nice-mc,dinh:2016density,rezende2015:variational,chen2018neural}
and they vary in expressiveness and computational cost.

In order to ensure that outputs of the network $\mu_{\alpha\beta}(\cdot)$
are in the correct well, a harmonic bias potential centered in the
target core is added during training
\begin{align}
V_{\mathrm{bias}}(\vec{x}) & =\begin{cases}
k\left(\vec{x}-\vec{x}_{\alpha}\right) & \vec{x}\in\Omega_{\alpha}\\
k\left(\vec{x}-\vec{x}_{\beta}\right) & \vec{x}\in\Omega_{\beta}
\end{cases},
\end{align}
where $\vec{x}_{\alpha}$ is the reference configuration in core $\alpha$,
resulting in the biased system $\tilde{V}(\vec{x})=V(\vec{x})+V_{\mathrm{bias}}(\vec{x})$
used during training. The network is trained in several stages, gradually
lowering the strength of the bias potential. To find the reference
configurations $\vec{x}_{\alpha}$ k-means clustering is run on samples
generated from local MCMC sampling in either well. Training sets of
both of the wells are generated for a set of gradually decreasing
bias strengths $\left\{ k_{i}\right\} _{i\leq N_{k}}$. After convergence
of the training at $k_{i},$ the training set is exchanged and training
is restarted with $k_{i+1}\leq k_{i}$. This allows for a slowly expanding
training set, which enables the network to learn how to generate meaningful
moves on a gradually more complex set of training data. The loss that
is to be minimized during training is given by the bi-directional
acceptance \ref{eq:loss_acceptance}

\begin{align}
C_{\mathrm{acc}} & =\mathbb{E}_{\vec{x}\sim\Omega_{\alpha}}\left\{ \left[\Delta\tilde{V}_{\alpha\beta}(\vec{x})+k_{B}T\log\left|\det J_{\mu_{\alpha\beta}}(\vec{x})\right|\right]^{2}\right\} ,\label{eq:neg_log_acceptance}
\end{align}
where the square of the norm is used in order to penalize high energies
more heavily. Training is performed in the forward and backward direction
and the same loss applies to samples from core $\Omega_{\beta}$ with
exchanged labels $\alpha\leftrightarrow\beta$.

\section{Numerical Experiments}

We demonstrate Neural MJMC on two examples: a two dimensional potential
landscape with three minima and a system consisting of two dimer particles
which are suspended in a bath of repulsive particles. The detailed
training parameters are given in Appendix C.

As a good compromise between computational cost and expressiveness,
we use real non-volume preserving transformations (RNVP) \cite{dinh:2016density}
for these examples. In a RNVP layer the configuration vector $\vec{x}\in\mathbb{R}^{N\times\mathrm{dim}}$
is split into two vectors $\vec{x}_{1}$ and $\vec{x}_{2}$. As we
deal with two dimensional systems, we split along the $x$ and $y$
coordinates of all particles such that $\vec{x}_{1},\vec{x}_{2}\in\mathbb{R}^{N\times\mathrm{1}}$.
One RNVP layer consists of two update steps in which the first subset
is updated based on the second while the second is kept constant,
and vice versa

\begin{align}
\left[\begin{array}{c}
\vec{x}_{1}^{(i)'}\\
\vec{x}_{2}^{(i)'}
\end{array}\right] & =\left[\begin{array}{c}
\vec{x}_{1}^{(i)}\odot\exp\left[S_{i}\left(\vec{x}_{2}^{(i)}\right)\right]+T_{i}\left(\vec{x}_{2}^{(i)}\right)\\
\vec{x}_{2}^{(i)}
\end{array}\right],\\
\left[\begin{array}{c}
\vec{x}_{1}^{(i+1)}\\
\vec{x}_{2}^{(i+1)}
\end{array}\right] & =\left[\begin{array}{c}
\vec{x_{1}}^{(i)'}\\
\vec{x}_{2}^{(i)'}\odot\exp\left[S_{i}'\left(\vec{x}_{1}^{(i)'}\right)\right]+T_{i}'\left(\vec{x}_{1}^{(i)'}\right)
\end{array}\right],
\end{align}

where the $S_{i},T_{i},T_{i}',S_{i}'$ are dense feed forward neural
networks. The above system of equations represent one RNVP block,
and arbitrary many of these blocks can be serially stacked resulting
in more complex transformations (see Fig. \ref{Fig:rnvp}). The logarithm
of the Jacobian determinant of this transformation is given by the
sum over all the outputs of all the scaling layers $\log\left|\det J_{\mu_{\alpha\beta}}(\vec{x})\right|=\sum_{i}\sum_{j}(S_{ij}+S_{ij}')$.

\begin{figure}
\centering\includegraphics[width=0.5\linewidth]{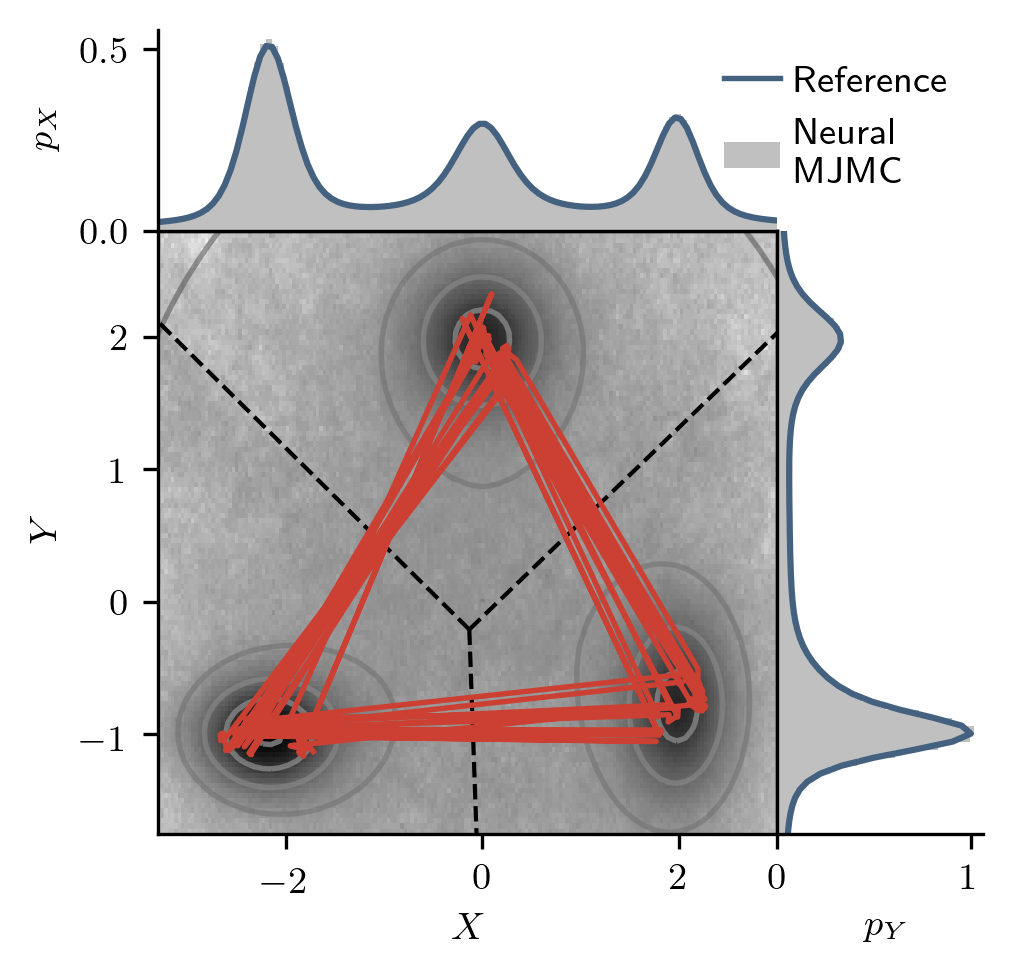}

\caption{Two dimensional histogram (center) of samples from the 2D Gaussian
triple well potential generated by Neural MJMC with a short section
of the Markov chain (red solid line) and marginal distributions $p_{X}$
(top) and $p_{Y}$ (right). The black dashed line depicts the border
between the states which are defined by a Voronoi tessellation. Convergence
to the correct Boltzmann distribution can be observed from the histograms
of the marginal distributions, where the blue line is the reference
solution from numerical integration of the system's Boltzmann distribution.}

\label{Fig:2D_Gaussian}
\end{figure}

\begin{figure}
\includegraphics{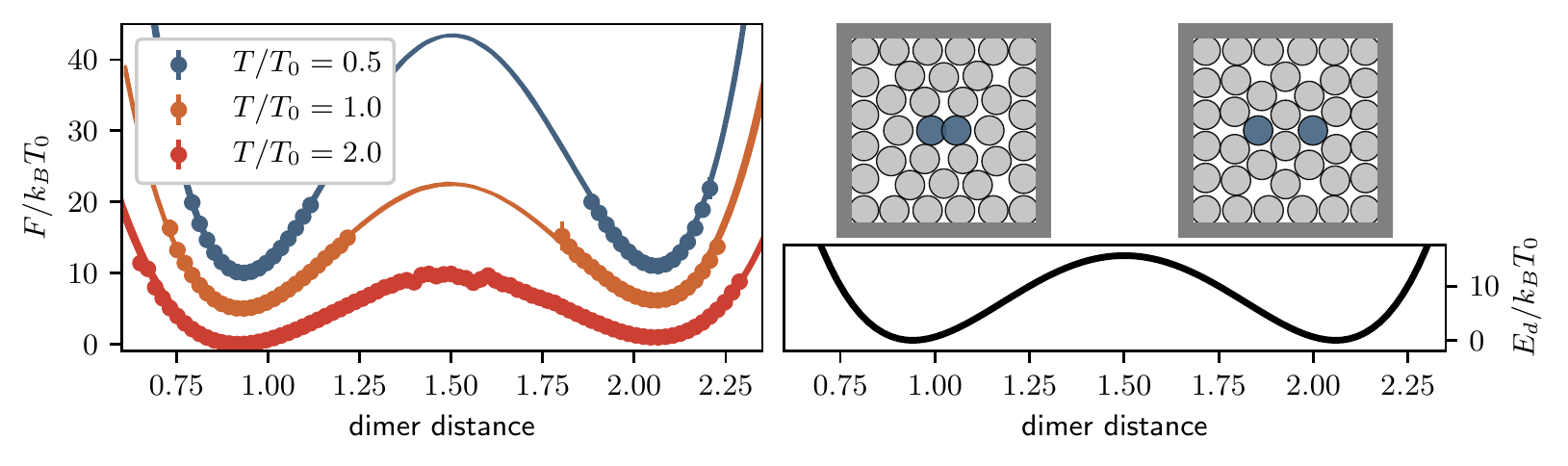}

\caption{\textbf{Left:} Free energy along the distance between the dimer particles.
The corresponding bands represent reference values obtained by umbrella
sampling. The neural network has been trained at temperature $T=T_{0}$,
then simulations at different temperatures have been performed using
Neural MJMC. Simulations are run for \SI{1.5e7}{} steps, and error
bars are generated from several sampling runs. In this figure, we
observe that Neural MJMC correctly samples the free energy along the
reaction coordinate of the system at different temperatures\textbf{.
Right top:} Reference configurations in the closed (left) and open
(right) dimer configuration. The dimer particles are displayed in
blue, and solvent particles in grey. The strongly repulsive potential
does not allow for significant overlaps between particles at equilibrium.\textbf{
Right bottom:} Dimer interaction potential $E_{d}$ as a function
of the dimer distance.}

\label{Fig:dimer_potential}
\end{figure}

\begin{figure}
\includegraphics{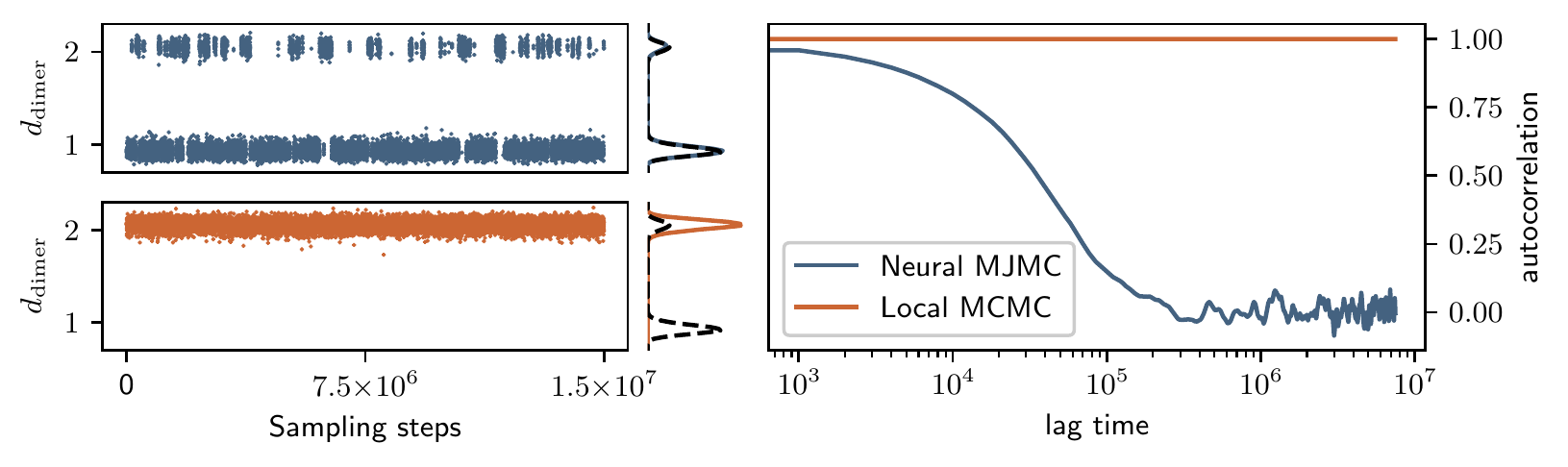}

\caption{\textbf{Left:} Dimer distance over a single realization using Neural
MJMC (top), and using local MCMC (bottom). Right) Histogram of the
dimer distance obtained by the displayed trajectory, with the reference
value displayed as the black dashed line. Spontaneous transitions
with local MCMC are not observed at this time scale. Neural MJMC explores
both metastable states in the trajectory multiple times and correctly
reproduces the distribution of dimer distances. \textbf{Right: }Autocorrelation
of the dimer distance. Neural moves allow for a fast exploration of
both metastable states, accelerating the production of uncorrelated
samples. In this figure, it is evident that Neural MJMC frequently
generates uncorrelated samples, and short trajectories are sufficient
to reconstruct the right distribution. In contrast, configurations
generated with local MCMC are highly correlated, as they do not cross
the energy barrier.}

\label{Fig:dimer_trajectories}
\end{figure}

\subsection{Gaussian triple well}

As an example for a system with multiple states, we demonstrate Neural
MJMC on a two dimensional potential landscape consisting of 3 Gaussian
shaped wells. We define the $3$ cores by a Voronoi tessellation for
which we use the minima of the Gaussians as centers. Each of the three
neural proposals are trained independently on configurations sampled
from the minima. In the sampling step, $100$ independent trajectories
of length \SI{e5}{\steps} are generated and averaged. We compare
the marginal distributions $p_{X}$ and $p_{Y}$ which are the projections
of the Boltzmann distribution on the $X$ and $Y$ axes and observe
great agreement to results from numerical integration of the Boltzmann
distribution (see Fig. \ref{Fig:2D_Gaussian}).

\subsection{Dimer in repulsive Lennard Jones bath}

As a bigger challenge Neural MJMC is applied to a two-dimensional
system composed of a bistable dimer immersed in a bath of strongly
repelling particles and confined to a box. The bistable dimer potential
has a minimum in the closed and open configurations which are separated
by a high energy barrier (see Fig. \ref{Fig:dimer_potential} right
bottom). Opening and closing of the dimer requires a concerted motion
of the solvent particles, that makes it difficult to sample the physical
path connecting the two configurations (see Ref. \cite{noe2019boltzmann}
for a more detailed description of the system).

The open and closed configuration serve as cores (see Fig. \ref{Fig:dimer_potential}
right top) in Neural MJMC and are distinguished by the distance between
the dimer particles. The neural network is trained on states sampled
independently in the closed and open configuration at four different
bias strengths with \SI{e5}{} samples for each well and bias. As
the system is invariant under permutation of solvent and dimer particles,
neural moves for each permutation of the system would have to be learned
independently. This is clearly unfeasible, as the number of permutation
scales factorial in the particle number. This is circumvented by permutation
reduction, i.e. re-labeling the particles such that the distance to
the reference configuration is minimized. This is realized using the
Hungarian algorithm \cite{hungarian} with the reference configurations
as target.

Each neural network in the RNVP architecture consists of three hidden
layer with $76$ nodes. The transformation consists of a total of
$20$ RNVP layers and contains approximately \SI{1.4e6}{} trainable
parameters. Neural MJMC is used to generate a single trajectory with
\SI{1.5e7}{\steps}, where the probability of neural moves is set
to \SI{1}{\percent}. In terms of computational performance, sampling
with Neural MJMC is approximately a factor of four slower than MCMC
with local displacements for this system. This slow down arises from
the evaluation of the network and the remapping of particles. As a
reference value, we use umbrella sampling to sample the free energy
along the dimer distance.

Neural moves cause direct transitions between the two metastable states
and thus a rapid exploration of the configuration space. The convergence
to the Boltzmann distribution is observed as shown in Fig. \ref{Fig:dimer_potential}.
An estimate for the crossing time with only local moves can be found
to be at the order of $10^{12}$ sampling steps at $T=1$ from the
Kramer's problem which make exhausting simulations unfeasible. In
Neural MJMC many crossings of the energy barrier can be observed (Fig.
\ref{Fig:dimer_trajectories} top). This is also reflected in the
autocorrelation function where samples generated with local MCMC remain
highly correlated, while it decays in Neural MJMC simulations on a
scale of approximately \SI{e5}{} sampling steps (see Fig. \ref{Fig:dimer_trajectories}
bottom). Thus generating the desired uncorrelated samples of the equilibrium
distribution.

\section{Conclusion}

In this paper, we have presented Neural Mode Jump Monte Carlo, a novel
method that allows for efficient sampling of Boltzmann distributions
of complex systems composed of many metastable states. The method
uses neural networks in order to parametrize bijections between metastable
regions in phase space and optimizes these networks for bi-directional
acceptance probability. By combining short steps given by random displacements
and large jumps between metastable states, the method is able to converge
quickly to the Boltzmann distribution. This is especially evident
in systems where large potential barriers are providing obstacles
to convergence of other methods. The method is demonstrated on two
toy examples, one with several bijections in two dimensions and a
high dimensional system consisting of a particle dimer in a bath of
Lennard Jones particles.

The reversible neural network architecture used in this work is also
used in the field of generative probabilistic modeling. Considering
the great attention this field is lately facing, it would not be surprising
to observe dramatic improvements in the performance of such reversible
networks. The efficiency and capability of Neural MJMC profoundly
rely on the specific architecture employed, and more sophisticated
networks would allow us to deal with systems of increasing complexity.
Neural MJMC is a general and transferable method and we can expect
it to be applied to a multitude of different systems.

\acks{We gratefully acknowledge funding from European Commission (ERC CoG 772230 "ScaleCell"), Deutsche Forschungsgemeinschaft (CRC1114/C03), the MATH+ Berlin Mathematics research center (AA1-6), and the International Max Planck Research School IMPRS-CBSC. Furthermore we want to thank Christoph Fröhner, Mohsen Sadeghi and Andreas Mardt for insightful discussions.}

\newpage{}

\begin{btSect}[plainnat]{literature}
\btPrintCited
\end{btSect}

\newpage{}

\appendix

\subsection*{Appendix A: Entropy difference \label{subsec:Appendix-A:-Entropy}}

The (Gibbs) entropy of a system is defined as
\begin{equation}
S_{X}=-k_{B}\int_{\Omega}p_{X}(x)\log p_{X}(x)dx.
\end{equation}

For a bijective function $y=\mu(x)$ we can apply the change of variable
formula to compute the change in entropy under the transformation.
With the transformed density being $p_{Y}(y)=p_{X}(\mu^{-1}(y))\left|\det J_{\mu^{-1}}(y)\right|$
we find

\begin{multline}
S_{X}=-k_{B}\int_{\mu(\Omega)}p_{Y}(y)\log p_{Y}(y)\left|\det J_{\mu^{=1}}(y)\right|dy\\
=S_{Y}-k_{B}\int_{\mu(\Omega)}p_{Y}(y)\log\left|\det J_{\mu^{-1}}(y)\right|dy.
\end{multline}

Thus the difference in entropy under the transformation $\mu(\cdot)$is
given as

\begin{equation}
\Delta S=S_{Y}-S_{X}=-k_{B}\mathbb{E}_{x\sim p_{\Omega}}\left[\log\left|\det J_{\mu}(x)\right|\right],
\end{equation}
where we used the inverse function theorem to compute the Jacobian.

\subsection*{Appendix B: Functional form of potentials}

Here we give the exact functional form of the potentials used to demonstrate
the proposed method.

\subsubsection*{Triple well potential}

Triple well potential is a 2D potential surface given by 
\begin{equation}
V(\vec{x})=\sum_{i}-a_{i}\exp\left[-\left(\vec{x}-\vec{m}_{i}\right)^{T}\Sigma_{i}\left(\vec{x}-\vec{m}_{i}\right)\right]+b\left\Vert \vec{x}\right\Vert ^{2},
\end{equation}
with $b=0.1$ and other parameters given in table \ref{table:parameters_triple_well}.
\begin{table}[h]
\caption{Parameters of the triple well potential}
\centering

\begin{tabular}{|c|c|c|c|}
\hline 
i & $\Sigma_{i}$ & $\vec{m}_{i}^{T}$ & $a_{i}$\tabularnewline
\hline 
\hline 
1 & $\textrm{diag}(0.5,0.3)$ & $(-2.2,-1)$ & $5$\tabularnewline
\hline 
2 & $\textrm{diag}(0.5,0.4)$ & $(0,2)$ & $5$\tabularnewline
\hline 
3 & $\textrm{diag}(0.4,0.5)$ & $(2,-0.8)$ & $5$\tabularnewline
\hline 
\end{tabular}\label{table:parameters_triple_well}
\end{table}

\subsubsection*{Dimer in a Lennard Jones bath}

The the dimer system is described in detail in Ref. \cite{noe2019boltzmann},
and the specific parameters used in this paper are given in table
\ref{table:dimer}.

\begin{table}
\caption{Parameters of the particle dimer system}
\centering

\begin{tabular}{|c|c|c|c|c|c|c|c|c|c|}
\hline 
Parameter & $\epsilon$ & $\sigma$ & $k_{d}$ & $d_{0}$ & $a$ & $b$ & $c$ & $l_{box}$ & $k_{box}$\tabularnewline
\hline 
\hline 
Value & 1.0 & 1.0 & 20 & 1.5 & 25 & 10 & 0.0 & 3.0 & 100\tabularnewline
\hline 
\end{tabular}

\label{table:dimer}
\end{table}

\subsection*{Appendix C: Details of the network architecture}

The RNVP network consists of may subsequent blocks which are depicted
in Fig. \ref{Fig:rnvp} (right). Each of these blocks consists of
$4$ independent networks, two for scaling and two for translation.
All networks use leaky ReLU in each hidden layer. The output of the
scaling networks uses a hyperbolic tangent scaled by a trainable scalar.
The output of the translation networks is linear. Adam \cite{kingma2014adam}
is used as optimizer with standard parameters and a learning rate
depending on the system. Table \ref{table:network_parameters} gives
an overview of the exact network architectures and hyperparameters
used in the experiments.
\begin{table}[h]
\caption{Parameters of the RNVP networks used in the experiments}
\centering

\begin{tabular}{|l|c|c|}
\hline 
 & DW & particles\tabularnewline
\hline 
\hline 
\# blocks & $10$ & $20$\tabularnewline
\hline 
hidden dimensions & $[20,20,20]$ & $[76,76,76]$\tabularnewline
\hline 
\# parameters & \SI{3.6e4}{} & \SI{1.4e6}{}\tabularnewline
\hline 
\# training samples per bias and core & \SI{1e5}{} & \SI{1e5}{}\tabularnewline
\hline 
bias strengths $/k_{B}T$ & $[10,0]$ & $[500,10,5,2]$\tabularnewline
\hline 
learning rate & \SI{e-3}{} & [\SI{e-3}{},\SI{e-4}{},\SI{e-4}{},\SI{e-5}{}]\tabularnewline
\hline 
batchsize & $2000$ & $8192$\tabularnewline
\hline 
\end{tabular}\label{table:network_parameters}
\end{table}

\ifanonsubmission
\else
\section*{Data and materials}
Computer code to generate the results presented in this paper is available at \href{https://github.com/noegroup/nmjmc}{here}.
\fi
\end{document}